# Subcarrier Domain of Multicarrier Continuous-Variable Quantum Key Distribution

Laszlo Gyongyosi

[1] Quantum Technologies Laboratory, Department of Telecommunications
*Budapest University of Technology and Economics*
2 Magyar tudosok krt, Budapest, *H*-1117, Hungary
[2] MTA-BME Information Systems Research Group
*Hungarian Academy of Sciences*
7 Nador st., Budapest, *H*-1051, Hungary

gyongyosi@hit.bme.hu

**Abstract**

We propose the subcarrier domain of multicarrier continuous-variable (CV) quantum key distribution (QKD). In a multicarrier CVQKD scheme, the information is granulated into Gaussian subcarrier CVs and the physical Gaussian link is divided into Gaussian sub-channels. The sub-channels are dedicated for the conveying of the subcarrier CVs. The angular domain utilizes the phase-space angles of the Gaussian subcarrier CVs to construct the physical model of a Gaussian sub-channel. The subcarrier domain injects physical attributes to the description of the subcarrier transmission. We prove that the subcarrier domain is a natural representation of the subcarrier-level transmission in a multicarrier CVQKD scheme. We also extend the subcarrier domain to a multiple-access multicarrier CVQKD setting. We demonstrate the results through the adaptive multicarrier quadrature-division (AMQD) CVQKD scheme and the AMQD-MQA (multiuser quadrature allocation) multiple-access multicarrier scheme. The subcarrier domain representation provides a general apparatus that can be utilized for an arbitrary multicarrier CVQKD scenario. The framework is particularly convenient for experimental multicarrier CVQKD scenarios.

# 1 Introduction

The continuous-variable quantum key distribution protocols allow for the parties to establish an unconditionally secure communication over the traditional telecommunication networks. In comparison to discrete-variable (DV) QKD, the CVQKD schemes do not require single-photon devices, a fact that allows us to implement in practice by standard devices [1–10], [24–31]. Despite the several attractive benefits of CVQKD, these protocols still require significant performance improvements to be comparable with that of the traditional telecommunications. For this purpose, the *multicarrier* CVQKD has been recently introduced through the adaptive multicarrier quadrature division (AMQD) modulation [2]. In a multicarrier CVQKD system, the Gaussian input CVs are granulated into subcarrier Gaussian CVs via the inverse Fourier transform, which are decoded by the receiver by the unitary CVQFT (continuous-variable quantum Fourier transform) operation [2]. Precisely, the multicarrier transmission divides the physical Gaussian link into several Gaussian sub-channels, where each sub-channel is dedicated for the conveying of a Gaussian subcarrier CV. In particular, the multicarrier transmission injects several benefits to CVQKD, such as improved secret key rates, higher tolerable excess noise, and enhanced transmission distances. Specifically, the benefits of the multicarrier CVQKD modulation has been extended to a multiple-access CVQKD through the AMQD-MQA (multiuser quadrature allocation) scheme [3], in which the simultaneous reliable transmission of the legal users is handled through the sophisticated allocation mechanism of the Gaussian subcarriers. The SVD-assisted (singular value decomposition) AMQD injects further extra degrees of freedom into the transmission [4], which can also be exploited in a multiple-access multicarrier CVQKD [5–7]. Here, we provide a natural representation of the subcarrier CV transmission and show that it also allows us to utilize the physical attributes into the sub-channel modeling.

The angular domain representation is a useful tool in traditional telecommunications to model the physical signal propagation through a communication channel. The angular domain representation is aimed at revealing and verifying the connections of the physical layer and the mathematical channel model in different scenarios, avoiding the use of an inaccurate channel representation [21–23]. Here, we show that similar benefits can be brought up to multicarrier CVQKD. We define the *subcarrier domain* representation for multicarrier CVQKD. The subcarrier domain utilizes the phase-space angles of the Gaussian subcarrier CV to construct the model of a Gaussian sub-channel and to build an appropriate statistical model of subcarrier transmission. The subcarrier domain is an adequate application for multicarrier CVQKD since it is a natural representation of the CVQFT operation. The key behind the subcarrier domain representation is the Fourier operation, which has a central role in a multicarrier CVQKD setting since this operation makes possible the construction of Gaussian subcarrier CVs from the single-carriers and Gaussian sub-channels from the physical Gaussian link.

Particularly, the CVQFT transformation not just opens the door for the characterization of the subcarrier domain of a Gaussian sub-channel but also provides us a framework to study the effects of psychical layer transmission in an experimental multicarrier CVQKD scenario. The subcarrier domain utilizes physical attributes such as the phase space angle into the description of the transmission. Thus, the subcarrier domain representation takes into account not just the theoretical model but also the physical level of the subcarrier transmission. Since the subcarrier domain is a natural representation of a multicarrier CVQKD transmission, it allows us to extend it to a multiple-access multicarrier CVQKD setting. Furthermore, the subcarrier domain model provides a general framework for any experiential multicarrier CVQKD.

This paper is organized as follows. In Section 2, some preliminaries are briefly summarized. Section 3 discusses the subcarrier domain representation for multicarrier CVQKD. Section 4 extends the subcarrier domain for multiple-access multiuser CVQKD. Finally, Section 5 concludes the results. Supplemental information is included in the Appendix.

## 2 Preliminaries

In Section 2, we briefly summarize the notations and basic terms. For further information, see the detailed descriptions of [2–6].

### 2.1 Basic Terms and Definitions

#### 2.1.1 Multicarrier CVQKD

In this section we very briefly summarize the basic notations of AMQD from [2]. The following description assumes a single user, and the use of $n$ Gaussian sub-channels $\mathcal{N}_i$ for the transmission of the subcarriers, from which only $l$ sub-channels will carry valuable information.

In the single-carrier modulation scheme, the $j$-th input single-carrier state $|\varphi_j\rangle = |x_j + \mathrm{i}p_j\rangle$ is a Gaussian state in the phase space $\mathcal{S}$, with i.i.d. Gaussian random position and momentum quadratures $x_j \in \mathbb{N}(0, \sigma_{\omega_0}^2)$, $p_j \in \mathbb{N}(0, \sigma_{\omega_0}^2)$, where $\sigma_{\omega_0}^2$ is the modulation variance of the quadratures. In the multicarrier scenario, the information is carried by Gaussian subcarrier CVs, $|\phi_i\rangle = |x_i + \mathrm{i}p_i\rangle$, $x_i \in \mathbb{N}(0, \sigma_\omega^2)$, $p_i \in \mathbb{N}(0, \sigma_\omega^2)$, where $\sigma_\omega^2$ is the modulation variance of the subcarrier quadratures, which are transmitted through a noisy Gaussian sub-channel $\mathcal{N}_i$. Precisely, each $\mathcal{N}_i$ Gaussian sub-channel is dedicated for the transmission of one Gaussian subcarrier CV from the $n$ subcarrier CVs. (*Note*: index $l$ refers to the subcarriers, while index $j$, to the single-carriers, throughout the manuscript.) The single-carrier state $|\varphi_j\rangle$ in the phase space $\mathcal{S}$ can be modeled as a zero-mean, circular symmetric complex Gaussian random variable $z_j \in \mathcal{CN}(0, \sigma_{\omega_{z_j}}^2)$, with variance $\sigma_{\omega_{z_j}}^2 = \mathbb{E}\left[|z_j|^2\right]$, and with i.i.d. real and imaginary zero-mean Gaussian random components $\mathrm{Re}(z_j) \in \mathbb{N}(0, \sigma_{\omega_0}^2)$, $\mathrm{Im}(z_j) \in \mathbb{N}(0, \sigma_{\omega_0}^2)$.

In the multicarrier CVQKD scenario, let $n$ be the number of Alice's input single-carrier Gaussian states. Precisely, the $n$ input coherent states are modeled by an $n$-dimensional, zero-mean, circular symmetric complex random Gaussian vector

$$\mathbf{z} = \mathbf{x} + \mathrm{i}\mathbf{p} = (z_1,\ldots,z_n)^T \in \mathcal{CN}(0,\mathbf{K_z}), \tag{1}$$

where each $z_j$ is a zero-mean, circular symmetric complex Gaussian random variable

$$z_j \in \mathcal{CN}\left(0,\sigma^2_{\omega_{z_j}}\right),\ z_j = x_j + \mathrm{i}p_j. \tag{2}$$

In the first step of AMQD, Alice applies the inverse FFT (fast Fourier transform) operation to vector $\mathbf{z}$ (see (1)), which results in an $n$-dimensional zero-mean, circular symmetric complex Gaussian random vector $\mathbf{d}$, $\mathbf{d} \in \mathcal{CN}(0,\mathbf{K_d})$, $\mathbf{d} = (d_1,\ldots,d_n)^T$, precisely as

$$\mathbf{d} = F^{-1}(\mathbf{z}) = e^{\frac{\mathbf{d}^T \mathbf{A} \mathbf{A}^T \mathbf{d}}{2}} = e^{\frac{\sigma^2_{\omega_0}(d_1^2 + \ldots + d_n^2)}{2}}, \tag{3}$$

where

$$d_i = x_{d_i} + \mathrm{i}p_{d_i},\ d_i \in \mathcal{CN}\left(0,\sigma^2_{d_i}\right), \tag{4}$$

where $\sigma^2_{\omega_{d_i}} = \mathbb{E}\left[|d_i|^2\right]$ and the position and momentum quadratures of $|\phi_i\rangle$ are i.i.d. Gaussian random variables

$$\mathrm{Re}(d_i) = x_{d_i} \in \mathbb{N}\left(0,\sigma^2_{\omega_i}\right),\ \mathrm{Im}(d_i) = p_{d_i} \in \mathbb{N}\left(0,\sigma^2_{\omega_i}\right), \tag{5}$$

where $\mathbf{K_d} = \mathbb{E}\left[\mathbf{d}\mathbf{d}^\dagger\right]$, $\mathbb{E}[\mathbf{d}] = \mathbb{E}\left[e^{\mathrm{i}\gamma}\mathbf{d}\right] = \mathbb{E}e^{\mathrm{i}\gamma}[\mathbf{d}]$, and $\mathbb{E}\left[\mathbf{d}\mathbf{d}^T\right] = \mathbb{E}\left[e^{\mathrm{i}\gamma}\mathbf{d}\left(e^{\mathrm{i}\gamma}\mathbf{d}\right)^T\right] = \mathbb{E}e^{\mathrm{i}2\gamma}\left[\mathbf{d}\mathbf{d}^T\right]$ for any $\gamma \in [0,2\pi]$. The $\mathbf{T}(\mathcal{N})$ transmittance vector of $\mathcal{N}$ in the multicarrier transmission is

$$\mathbf{T}(\mathcal{N}) = [T_1(\mathcal{N}_1),\ldots,T_n(\mathcal{N}_n)]^T \in \mathcal{C}^n, \tag{6}$$

where

$$T_i(\mathcal{N}_i) = \mathrm{Re}(T_i(\mathcal{N}_i)) + \mathrm{i}\,\mathrm{Im}(T_i(\mathcal{N}_i)) \in \mathcal{C}, \tag{7}$$

is a complex variable, which quantifies the position and momentum quadrature transmission (i.e., gain) of the $i$-th Gaussian sub-channel $\mathcal{N}_i$, in the phase space $\mathcal{S}$, with real and imaginary parts

$$0 \leq \mathrm{Re}\,T_i(\mathcal{N}_i) \leq 1/\sqrt{2},\ \text{and}\ 0 \leq \mathrm{Im}\,T_i(\mathcal{N}_i) \leq 1/\sqrt{2}. \tag{8}$$

Particularly, the $T_i(\mathcal{N}_i)$ variable has the squared magnitude of

$$|T_i(\mathcal{N}_i)|^2 = \mathrm{Re}\,T_i(\mathcal{N}_i)^2 + \mathrm{Im}\,T_i(\mathcal{N}_i)^2 \in \mathbb{R}, \tag{9}$$

where

$$\mathrm{Re}\,T_i(\mathcal{N}_i) = \mathrm{Im}\,T_i(\mathcal{N}_i). \tag{10}$$

The Fourier-transformed transmittance of the $i$-th sub-channel $\mathcal{N}_i$ (resulted from CVQFT operation at Bob) is denoted by

$$|F(T_i(\mathcal{N}_i))|^2. \tag{11}$$

The $n$-dimensional zero-mean, circular symmetric complex Gaussian noise vector $\Delta \in \mathcal{CN}\left(0, \sigma_\Delta^2\right)_n$ of the quantum channel $\mathcal{N}$, is evaluated as

$$\Delta = \left(\Delta_1, \ldots, \Delta_n\right)^T \in \mathcal{CN}\left(0, \mathbf{K}_\Delta\right), \tag{12}$$

where

$$\mathbf{K}_\Delta = \mathbb{E}\left[\Delta \Delta^\dagger\right], \tag{13}$$

with independent, zero-mean Gaussian random components

$$\Delta_{x_i} \in \mathbb{N}\left(0, \sigma^2_{\mathcal{N}_i}\right), \text{ and } \Delta_{p_i} \in \mathbb{N}\left(0, \sigma^2_{\mathcal{N}_i}\right), \tag{14}$$

with variance $\sigma^2_{\mathcal{N}_i}$, for each $\Delta_i$ of a Gaussian sub-channel $\mathcal{N}_i$, which identifies the Gaussian noise of the $i$-th sub-channel $\mathcal{N}_i$ on the quadrature components in the phase space $\mathcal{S}$.

The CVQFT-transformed noise vector can be rewritten as

$$F(\Delta) = \left(F(\Delta_1), \ldots, F(\Delta_n)\right)^T, \tag{15}$$

with independent components $F(\Delta_{x_i}) \in \mathbb{N}\left(0, \sigma^2_{F(\mathcal{N}_i)}\right)$ and $F(\Delta_{p_i}) \in \mathbb{N}\left(0, \sigma^2_{F(\mathcal{N}_i)}\right)$ on the quadratures, for each $F(\Delta_i)$. Precisely, it also defines an $n$-dimensional zero-mean, circular symmetric complex Gaussian random vector $F(\Delta) \in \mathcal{CN}\left(0, \mathbf{K}_{F(\Delta)}\right)$ with a covariance matrix

$$\mathbf{K}_{F(\Delta)} = \mathbb{E}\left[F(\Delta) F(\Delta)^\dagger\right]. \tag{16}$$

## 3 Subcarrier Domain of Multicarrier CVQKD

**Proposition 1** (Subcarrier domain representation of multicarrier transmission.) *For the $i$-th Gaussian sub-channel $\mathcal{N}_i$, the subcarrier domain representation is $\mathcal{R}_{\phi_i}\left(T_i(\mathcal{N}_i)\right) = U T_i(\mathcal{N}_i) U$, where $U$ is the CVQFT operator.*

*Proof.*
The proofs throughout assume $l$ Gaussian sub-channels for the multicarrier transmission. The angles of the $|\phi_i\rangle$ transmitted and the $|\phi_i'\rangle$ received subcarrier CVs in the phase space $\mathcal{S}$ are denoted by $\theta_i^* \in [0, 2\pi]$, and $\theta_i \in [0, 2\pi]$, respectively.

Specifically, first, we express $F(\mathbf{T}(\mathcal{N}))$ as

$$\begin{aligned} F(\mathbf{T}(\mathcal{N})) &= \sum_l F(T_i(\mathcal{N}_i)) \\ &= \sum_{i=0}^{l-1} \sum_{k=0}^{l-1} T_k e^{\frac{-i2\pi ik}{l}}, \end{aligned} \tag{17}$$

from which $F(T_i(\mathcal{N}_i))$ is yielded as

$$F\left(T_i\left(\mathcal{N}_i\right)\right) = \sum_{k=0}^{l-1} T_k e^{\frac{-i2\pi ik}{l}}. \tag{18}$$

Next, we recall the attributes of a multicarrier CVQKD transmission from [2]. In particular, assuming $l$ Gaussian sub-channels, the output $\mathbf{y}$ is precisely as follows:

$$\begin{aligned} \mathbf{y} &= F\left(\mathbf{T}\left(\mathcal{N}\right)\right) F\left(\mathbf{d}\right) + F\left(\Delta\right) \\ &= F\left(\mathbf{T}\left(\mathcal{N}\right)\right) F\left(F^{-1}\left(\mathbf{z}\right)\right) + F\left(\Delta\right) \\ &= \left(F\left(\mathbf{T}\left(\mathcal{N}\right)\right) F\right) \mathbf{d} + F\left(\Delta\right) \\ &= \sum_l F\left(T_i\left(\mathcal{N}_i\right)\right) d_i + F\left(\Delta_i\right), \end{aligned} \tag{19}$$

where

$$F\left(\mathbf{d}\right) = F\left(F^{-1}\left(\mathbf{z}\right)\right) = \mathbf{z}. \tag{20}$$

The $l$ columns of the $l \times l$ unitary matrix $F$ formulate basis vectors, which are referred to as the domain $\mathcal{R}_{\phi_i}$, from which the subcarrier domain representation $\mathcal{R}_{\phi_i}\left(T_i\left(\mathcal{N}_i\right)\right)$ of $T_i\left(\mathcal{N}_i\right)$ is defined as

$$\mathcal{R}_{\phi_i}\left(T_i\left(\mathcal{N}_i\right)\right) = F\left(T_i\left(\mathcal{N}_i\right)\right) F, \tag{21}$$

Thus, (19) can be rewritten as

$$\begin{aligned} \mathbf{y}^{\mathcal{R}_\phi} &= \mathcal{R}_\phi\left(\mathbf{T}\left(\mathcal{N}\right)\right) \mathbf{d} + F\left(\Delta\right) \\ &= \sum_l \mathcal{R}_{\phi_i}\left(T_i\left(\mathcal{N}_i\right)\right) d_i + F\left(\Delta_i\right), \end{aligned} \tag{22}$$

where $\mathbf{y}^{\mathcal{R}_\phi}$ is referred to as the *subcarrier domain representation* of $\mathbf{y}$.

Particularly, from (21) follows that $\mathcal{R}_{\phi_i}\left(T_i\left(\mathcal{N}_i\right)\right)$ can be expressed as

$$\mathcal{R}_{\phi_i}\left(T_i\left(\mathcal{N}_i\right)\right) = U T_i\left(\mathcal{N}_i\right) U, \tag{23}$$

where $U$ is an $l \times l$ unitary matrix as

$$U = F, \tag{24}$$

where $F$ refers to the CVQFT operator which for $l$ subcarriers can be expressed by an $l \times l$ matrix, as

$$U = \tfrac{1}{\sqrt{l}} e^{\frac{-i2\pi ik}{l}}, \; i,k = 0,\ldots,l-1, \tag{25}$$

Thus,

$$\mathcal{R}_{\phi_i}\left(\mathbf{T}\left(\mathcal{N}\right)\right) = \sum_{i=1}^{l} U\left(T_i\left(\mathcal{N}_i\right)\right). \tag{26}$$

To conclude, the results in (17) and (18) can be rewritten as

$$U\left(\mathbf{T}\left(\mathcal{N}\right)\right) = \sum_{i=1}^{l} U\left(T_i\left(\mathcal{N}_i\right)\right) \tag{27}$$

thus,

$$F\left(T_i\left(\mathcal{N}_i\right)\right) = U\left(T_i\left(\mathcal{N}_i\right)\right). \tag{28}$$

Specifically, an arbitrary distributed $\mathcal{R}_\phi\left(\mathbf{T}\left(\mathcal{N}\right)\right)$ can be approximated via an averaging over the following statistics:

$$\mathcal{S}\big(\mathcal{R}_\phi\big(\mathbf{T}(\mathcal{N})\big)\big) \in \mathcal{CN}\big(0, \sigma^2_{\mathbf{T}(\mathcal{N})}\big), \tag{29}$$

by theory.

Since the unitary $U$ operation does not change the distribution of $\mathcal{S}(\mathbf{T}(\mathcal{N}))$, an arbitrarily distributed $\mathbf{T}(\mathcal{N})$ can be approximated via an averaging over the statistics of

$$\mathcal{S}\big(\mathbf{T}(\mathcal{N})\big) \in \mathcal{CN}\big(0, \sigma^2_{\mathbf{T}(\mathcal{N})}\big). \tag{30}$$

∎

**Theorem 1** (Subcarrier domain of a Gaussian sub-channel). *The $\mathcal{R}_{\phi_i}$ subcarrier domain representation of $\mathcal{N}_i$, $i = 0...l-1$, is $\mathcal{R}_{\phi_i}(T_i(\mathcal{N}_i)) = \sum_k \mathrm{A}(\mathcal{N}_i) b(k/l)^\dagger b(\cos\theta_i) b(\cos\theta_i^*)^\dagger b(i/l)$, $k = 0...l-1$, where $b(\cdot)$ is an orthonormal basis vector of $\mathcal{R}_{\phi_i}$, $\theta_i^*$ and $\theta_i$ are the phase-space angles of $|\phi_i\rangle$ and $|\phi_i'\rangle$, $\mathrm{A}(\mathcal{N}_i) = x_i$, where $x_i$ is a real variable, $x_i \geq 0$.*

*Proof.*

The $b$ basis vectors of $\mathcal{R}_{\phi_i}$ are evaluated as follows: Let $\theta_i^* \in [0, 2\pi]$ refer to the angle of the $i$-th noise-free input Gaussian subcarrier CV $|\phi_i\rangle$ in $\mathcal{S}$. The angle of the $i$-th noisy subcarrier CV $|\phi_i'\rangle$ is referred to as $\theta_i \in [0, 2\pi]$, $\theta_i \neq \theta_i^*$.

In particular, for the subcarrier domain representation, the scaled CVQFT operation defines the $b(\cdot)$ basis at $l$ Gaussian sub-channels as an $l \times 1$ matrix:

$$b(\cos\theta_i) = \frac{1}{\sqrt{l}} \begin{pmatrix} 1 \\ e^{-\mathrm{i}2\pi\cos\theta_i} \\ e^{-\mathrm{i}2\pi 2 \cos\theta_i} \\ \vdots \\ e^{-\mathrm{i}2\pi(l-1)\cos\theta_i} \end{pmatrix}. \tag{31}$$

while for the input angle $\theta_i^*$ also defines an $l \times 1$ matrix as

$$b(\cos\theta_i^*) = \frac{1}{\sqrt{l}} \begin{pmatrix} 1 \\ e^{-\mathrm{i}2\pi\cos\theta_i^*} \\ e^{-\mathrm{i}2\pi 2 \cos\theta_i^*} \\ \vdots \\ e^{-\mathrm{i}2\pi(l-1)\cos\theta_i^*} \end{pmatrix}. \tag{32}$$

Precisely, the difference of the cos functions of the $i$-th $\theta_i^*$ transmitted and the $\theta_i$ received angles is defined as

$$\tau_i = \cos\theta_i - \cos\theta_i^*. \tag{33}$$

Let $b(\cos\theta_i)$ and $b(\cos\theta_i^*)$ be the basis vectors of $\cos\theta_i, \cos\theta_i^*$ [21–23], then for the $\Omega_i = \theta_i - \theta_i^*$ angle

$$|\cos\Omega_i| = \left|b(\cos\theta_i^*)^\dagger b(\cos\theta_i)\right|, \quad (34)$$

and

$$|\cos\Omega_i| = |f(\tau_i)|, \quad (35)$$

where [21]

$$f(\tau_i) = \frac{1}{l} e^{i\pi(l-1)(\cos\theta_i - \cos\theta_i^*)} \frac{\sin(\pi l(\cos\theta_i - \cos\theta_i^*))}{\sin(\pi(\cos\theta_i - \cos\theta_i^*))}. \quad (36)$$

In particular, using (36), after some calculations, the result in (34) can be rewritten as

$$|\cos\Omega_i| = \left|\frac{\sin(\pi l(\cos\theta_i - \cos\theta_i^*))}{l\sin(\pi(\cos\theta_i - \cos\theta_i^*))}\right|. \quad (37)$$

Specifically, by expressing (36) via the formula of

$$\operatorname{sinc}(x) = \sin(\pi x)\frac{1}{\pi x}, \quad (38)$$

one can find that for $l \to \infty$, $f(\tau_i)$ can be rewritten as [21]

$$\lim_{l\to\infty} f(\tau_i) = e^{i\pi l \tau_i}\operatorname{sinc}(l\tau_i). \quad (39)$$

The function $|f(\tau_i)|$ for different values of $\tau_i$ is depicted in Fig. 1. The function picks up the $|f(\tau_i)| = 1$ maximum at $\tau_i = 0$, with a period $r = 1$. For $l$ sub-channels, a period yields $l$ values.

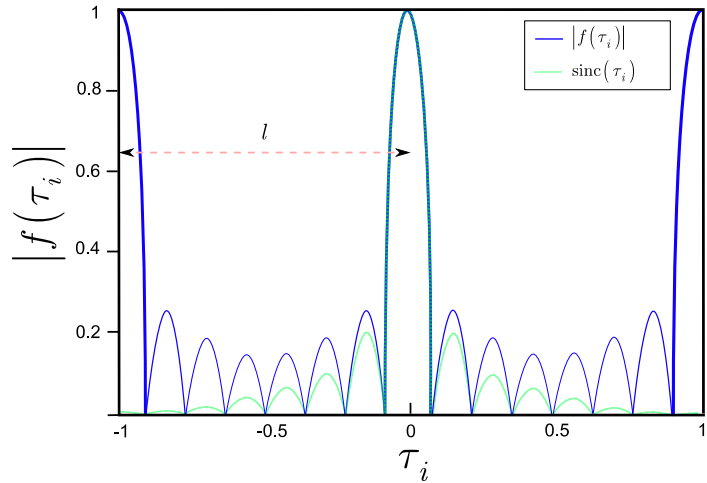

**Figure 1.** The function $|f(\tau_i)|$ (blue) for different values of $\tau_i$. The period of the function is $r = 1$. The sinc function (green) is approximated with an arbitrary precision in the asymptotic limit of $l \to \infty$.

Next, we utilize function $f(\cdot)$ to derive the $\mathcal{R}_{\phi_i}$ subcarrier domain representation of $\mathcal{N}_i$. Function $f(\tau_i)$ at a given $\theta_i$ formulates a plot

$$\mathrm{p}:\left(\cos\theta_i,\left|f\left(\tau_i\right)\right|\right), \tag{40}$$

where from the $r = 1$ periodicity of $f(\cdot)$ follows that the main loops are obtained at

$$\cos\theta_i = \cos\theta_i^*. \tag{41}$$

The plot $\left|f\left(\tau_i\right)\right|$ as a function of $\cos\theta_i$ is depicted in Fig. 2, for $\theta_i^* = \pi/2$, $l = 2$.

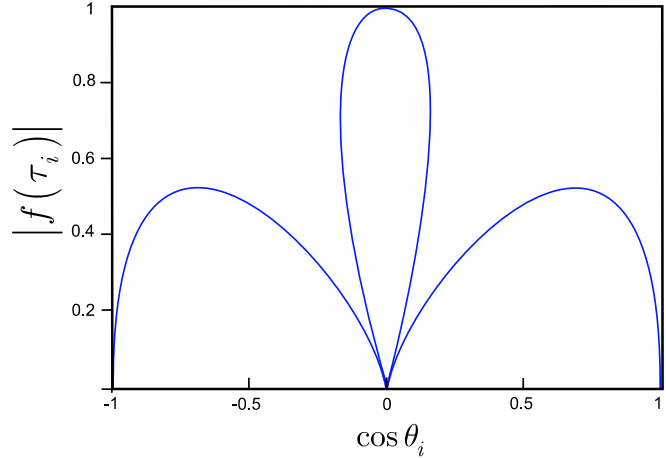

**Figure 2.** The $\mathcal{R}_{\phi_i}$ subcarrier domain representation of $\mathcal{N}_i$, at $\theta_i^* = \pi/2$, $l = 2$. The function $\left|f\left(\tau_i\right)\right|$ picks up the maximum at $\tau_i = 0$.

From the bases $b(\cdot)$ of (31) and (32), the Fourier bases $b\left(\frac{k}{l}\right)$ and $b\left(\frac{i}{l}\right)$, $i,k = 0,\ldots,l-1$ is defined as follows:

The set $\mathrm{S}_b$ of the $\mathcal{R}_\phi$ orthonormal basis over the $\mathcal{C}^l$ complex space of the $\mathcal{R}_\phi$ subcarrier domain representation can be defined as

$$\mathrm{S}_b = \left\{b(0), b\left(\tfrac{1}{l}\right), \ldots, b\left(\tfrac{l-1}{l}\right)\right\} \in \mathcal{C}^l, \tag{42}$$

and $b\left(\frac{k}{l}\right)$ is an $l \times 1$ matrix as

$$b(0) = \tfrac{1}{\sqrt{l}}\begin{pmatrix}1\\1\\1\\\vdots\\1\end{pmatrix},\text{ and } b\left(\tfrac{k}{l}\right) = \tfrac{1}{\sqrt{l}}\begin{pmatrix}1\\e^{\frac{-i2\pi k}{l}}\\e^{\frac{-i2\pi 2k}{l}}\\\vdots\\e^{\frac{-i2\pi(l-1)k}{l}}\end{pmatrix}, \tag{43}$$

while the $b\left(\frac{i}{l}\right)$ $l \times 1$ matrix is precisely as

$$b\left(\tfrac{i}{l}\right) = \tfrac{1}{\sqrt{l}} \begin{pmatrix} 1 \\ e^{\frac{-i2\pi i}{l}} \\ e^{\frac{-i2\pi i 2}{l}} \\ \vdots \\ e^{\frac{-i2\pi i(l-1)}{l}} \end{pmatrix}. \tag{44}$$

Precisely, using the orthonormal basis of (42), the result in (36) can be rewritten as

$$f(\tau_i) = b(0)^\dagger b(\tau_i). \tag{45}$$

Specifically, the expression of (45) allows us to redefine the plot of (40) to express $b\left(\tfrac{k}{l}\right)$ as follows:

$$\mathrm{p}_{b\left(\tfrac{k}{l}\right)} : \left(\cos\theta_i, \left|f\left(\cos\theta_i - \tfrac{k}{l}\right)\right|\right), \tag{46}$$

and thus the maximum values are obtained at

$$\cos\theta_i = \tfrac{k}{l}. \tag{47}$$

Particularly, at a given $l$, evaluating $f$ at $k = 1,\ldots,l-1$ yields the following values [21]:

$$f\left(\tfrac{k}{l}\right) = 0, \tag{48}$$

and

$$f\left(\tfrac{-k}{l}\right) = f\left(\tfrac{l-k}{l}\right). \tag{49}$$

In particular, the $A(\mathcal{N}_i)$ parameter is called the virtual gain of the $\mathcal{N}_i$ sub-channel transmittance coefficient, and without loss of generality, it is defined as

$$A(\mathcal{N}_i) = x_i, \tag{50}$$

where $x_i$ is a real variable, $x_i \geq 0$.

From (31), (32), and (50), $\mathbf{T}(\mathcal{N})$ can be expressed as

$$\mathbf{T}(\mathcal{N}) = \sum_{i=0}^{l-1} A(\mathcal{N}_i) b(\cos\theta_i) b(\cos\theta_i^*)^\dagger. \tag{51}$$

By exploiting the properties of the Fourier transform [21–23], for a given $\cos\theta_i^*$ and $\cos\theta_i$, $\mathcal{R}_\phi(\mathbf{T}(\mathcal{N}))$ can be rewritten as

$$\begin{aligned}
\mathcal{R}_\phi(\mathbf{T}(\mathcal{N})) &= \sum_{i=0}^{l-1}\sum_{k=0}^{l-1} b\left(\tfrac{k}{l}\right)^\dagger T_i(\mathcal{N}_i) b\left(\tfrac{i}{l}\right) \\
&= \sum_{i=0}^{l-1}\sum_{k=0}^{l-1} b\left(\tfrac{k}{l}\right)^\dagger A(\mathcal{N}_i) b(\cos\theta_i) b(\cos\theta_i^*)^\dagger b\left(\tfrac{i}{l}\right) \\
&= \sum_{i=0}^{l-1}\sum_{k=0}^{l-1} x_i b\left(\tfrac{k}{l}\right)^\dagger b(\cos\theta_i) b(\cos\theta_i^*)^\dagger b\left(\tfrac{i}{l}\right).
\end{aligned} \tag{52}$$

Specifically, in (52), for the representation of term $b(\cos\theta_i^*)^\dagger b(i/l)$ [21], set $S^{\mathcal{R}_{\phi_i}}$ can be defined for the $\mathcal{R}_{\phi_i}$ domain of $\mathcal{N}_i$ as

$$S^{\mathcal{R}_\phi} : \left|\cos\theta_i - \left(\tfrac{k}{l}\right)\right| < \tfrac{1}{l}. \tag{53}$$

The set $S^{\mathcal{R}_\phi}$ in the subcarrier domain representation for $\theta_i^* = \pi/2$, $l=2$ and $k=0,\ldots,l-1$ is illustrated by the dashed area in Fig. 3. The $b(0), b\left(\tfrac{1}{l}\right),\ldots,b\left(\tfrac{l-k}{l}\right)$ basis vectors of $\mathcal{R}_\phi$ for $l=2$ are also depicted, evaluated via (46).

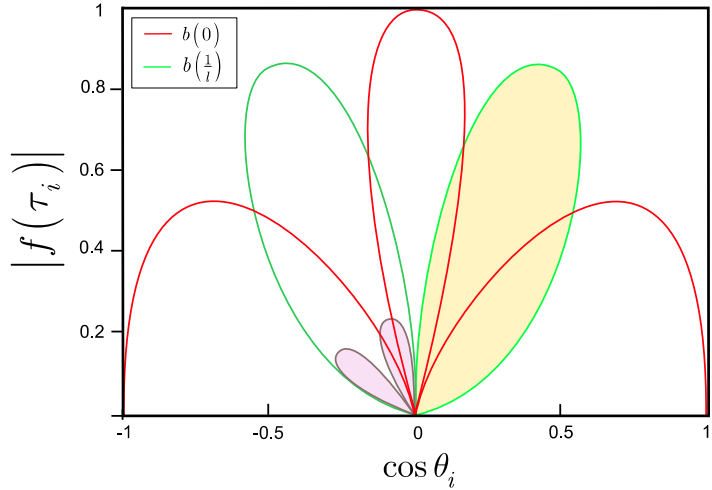

**Figure 3**. The set $S^{\mathcal{R}_\phi}$ (dashed areas) for $\theta_i^* = \pi/2$, $l=2$ Gaussian sub-channels and $k=0,\ldots l-1$. The curves (red, green) depict the basis vectors $b(0), b\left(\tfrac{1}{l}\right)$ of $\mathcal{R}_\phi$.

Putting the pieces together, from (52), $\mathcal{R}_{\phi_i}\left(T_i(\mathcal{N}_i)\right)$ of a given Gaussian sub-channel $\mathcal{N}_i$ is yielded as follows:

$$\begin{aligned}
&\mathcal{R}_{\phi_i}\left(T_i(\mathcal{N}_i)\right) \\
&= \sum_{k=0}^{l-1} b\left(\tfrac{k}{l}\right)^\dagger T_i(\mathcal{N}_i) b\left(\tfrac{i}{l}\right) \\
&= \sum_{k=0}^{l-1} \mathrm{A}(\mathcal{N}_i) b\left(\tfrac{k}{l}\right)^\dagger b\left(\cos\theta_i\right) b\left(\cos\theta_i^*\right)^\dagger b\left(\tfrac{i}{l}\right).
\end{aligned} \tag{54}$$

∎

## 3.1 Statistics of Subcarrier Domain Sub-channel Transmission

**Theorem 2** (Transmittance of the Gaussian sub-channels). *For arbitrary $\theta_i^*$, the magnitude $\left|\mathcal{R}_{\phi_i}\left(T_i(\mathcal{N}_i)\right)\right|$ of $\mathcal{R}_{\phi_i}\left(T_i(\mathcal{N}_i)\right)$ of a given $\mathcal{N}_i$ is maximized in the asymptotic limit of $\cos\Omega_i \to 1$, where $\Omega_i = \theta_i - \theta_i^*$. Averaging over the statistics $\mathcal{S}\left(\mathcal{R}_\phi(\mathbf{T}(\mathcal{N}))\right) \in \mathcal{CN}\left(0,\sigma^2_{\mathbf{T}(\mathcal{N})}\right)$, $\mathrm{rank}\left(\mathcal{S}\left(\mathcal{R}_\phi(\mathbf{T}(\mathcal{N}))\right)\right) = \min\left(|S_i|,|S_k|\right)$, where the cardinality of sets $S_i, S_k$ identifies the number of non-zero rows and columns of $\mathcal{R}_{\phi_i}(\mathbf{T}(\mathcal{N}))$.*

*Proof.*

First, we recall $\mathcal{R}_{\phi_i}(T_i(\mathcal{N}_i))$ from (54) and express $\left|\mathcal{R}_{\phi_i}(T_i(\mathcal{N}_i))\right|$ as

$$\left|\mathcal{R}_{\phi_i}(T_i(\mathcal{N}_i))\right| = \left|\sum_{k=0}^{l-1} x_i b\left(\tfrac{k}{l}\right)^{\dagger} b(\cos\theta_i) b(\cos\theta_i^*)^{\dagger} b\left(\tfrac{i}{l}\right)\right|. \tag{55}$$

For a given $\theta_i^*$, the values of angle $\theta_i$ has the following statistical impacts on (55).

Without loss of generality, let parameters $i$ and $k$ be fixed as

$$i, k = \{C, C\}, \tag{56}$$

where $C > 0$ is a real variable.

Particularly, for the $\mathcal{N}_i$ sub-channels, the $\left|\mathcal{R}_{\phi_i}(T_i(\mathcal{N}_i))\right|$ magnitudes formulate a set

$$\partial = \left\{\left|\mathcal{R}_{\phi_i}(T_i(\mathcal{N}_i))\right|, i = 0, \ldots, l-1\right\}. \tag{57}$$

Let

$$\Omega_i = \theta_i - \theta_i^*, \tag{58}$$

and let $s$ be the number of sub-channels for which $\left|\mathcal{R}_{\phi_i}(T_i(\mathcal{N}_i))\right| \approx 0$.

Specifically, for $\partial$, let determine $\Omega_i$ the value of $k$ as

$$k = \begin{cases} k \in [0, 2C], & \text{if } |\Omega_i| \to \pi \\ k = i = C, & \text{if } |\Omega_i| = 0 \end{cases}. \tag{59}$$

Let us define a $\mathcal{G}_0$ initial subset with $|\mathcal{G}_0| = s_0$, as

$$\mathcal{G}_0 = \left\{\left|\mathcal{R}_{\phi_j}(T_j(\mathcal{N}_j))\right|, j = 0, \ldots, s_0 - 1\right\} \subseteq \partial, \tag{60}$$

where

$$\left|\mathcal{R}_{\phi_j}(T_j(\mathcal{N}_j))\right| \approx 0. \tag{61}$$

In this setting, as $\cos\Omega_i \to 1$, the cardinality of $\mathcal{G}$ increases,

$$\mathcal{G} : \cos\Omega_i \to 1 : |\mathcal{G}| = s > |\mathcal{G}_0| = s_0, \tag{62}$$

while as $\cos\Omega_i \to -1$, the cardinality of $\mathcal{G}$ decreases, thus

$$\mathcal{G} : \cos\Omega_i \to -1 : |\mathcal{G}| = s < |\mathcal{G}_0| = s_0. \tag{63}$$

In particular, as (63) holds, the range of $k$ expands from $C$ to the full domain of $k = [0, 2C]$, and $\left|\mathcal{R}_{\phi_i}(T_i(\mathcal{N}_i))\right|$ decreases, thus

$$\left|\mathcal{R}_{\phi_i}(T_i(\mathcal{N}_i))\right| \approx a, \tag{64}$$

where $a$ is an average which around the $\left|\mathcal{R}_{\phi_i}(T_i(\mathcal{N}_i))\right|$ coefficients stochastically moves [21–23]. The impact of $\cos\Omega_i \to -1$ on $\left|\mathcal{R}_{\phi_i}(T_i(\mathcal{N}_i))\right|$ is depicted in Fig. 4 for $i, k = \{C, C\}$, $i = C = 0.5l$. The maximum transmittance is normalized to unit for $k = 0.5C$. Statistically, the convergence of $\cos\Omega_i \to -1$ improves the range of $k$ and decreases the sub-channel transmittance (see (59)).

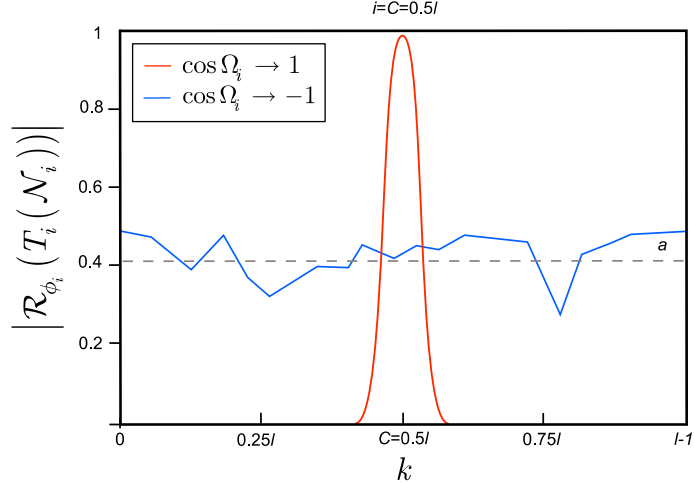

**Figure 4**. The impact of $\cos\Omega_i \to -1$ on $\left|\mathcal{R}_{\phi_i}\left(T_i\left(\mathcal{N}_i\right)\right)\right|$ for a fixed $i = C = 0.5l$ on a sub-channel $\mathcal{N}_i$. The parameter range changes to $i,k = \{C, [0, 2C]\}$, $C = 0.5l$, from $i,k = \{C,C\}$. For $\cos\Omega_i \to 1$, the transmittance picks up the maximum at $k = C = 0.5l$ (red) in a narrow range of $k \approx C$. Statistically, as $\cos\Omega_i \to -1$, the transmittance significantly decreases, moving stochastically around an average $a$ (dashed grey line) within the full range $k = [0, 2C]$.

Particularly, the degrees of freedom in $\mathcal{R}_\phi(\mathbf{T}(\mathcal{N}))$ can be evaluated through the rank of $\mathcal{R}_\phi(\mathbf{T}(\mathcal{N}))$.

Let us identify the number of non-zero rows and columns of $\mathcal{R}_{\phi_i}(\mathbf{T}(\mathcal{N}))$ via $|S_i|$ and $|S_k|$, of sets $S_i, S_k$, respectively. By averaging [21–23] over the statistics of
$$\mathcal{S}\left(\mathcal{R}_\phi\left(\mathbf{T}(\mathcal{N})\right)\right) \in \mathcal{CN}\left(0, \sigma^2_{\mathbf{T}(\mathcal{N})}\right), \tag{65}$$
thus the rank of $\mathcal{S}\left(\mathcal{R}_\phi\left(\mathbf{T}(\mathcal{N})\right)\right)$ without loss of generality is expressed as
$$\begin{aligned} rank\left(\mathcal{S}\left(\mathcal{R}_\phi\left(\mathbf{T}(\mathcal{N})\right)\right)\right) &= \min\left(|S_i|, |S_k|\right) \\ &\approx \min\left(\sum_l \cos\theta_i, \sum_l \cos\theta_i^*\right), \end{aligned} \tag{66}$$
for an arbitrary distribution, by theory [21].

The rank in (66) basically changes in function of the number $l$ of $\mathcal{N}_i$ Gaussian sub-channels utilized for the multicarrier transmission since the increasing $l$ results in more non-zero elements in $\mathcal{S}\left(\mathcal{R}_\phi\left(\mathbf{T}(\mathcal{N})\right)\right)$ [21–23]. On the other hand, the rank in (66) also changes in function of $\theta_i$. Specifically, as $\cos\Omega_i \to 1$, the matrix $\mathcal{S}\left(\mathcal{R}_\phi\left(\mathbf{T}(\mathcal{N})\right)\right)$ will have significantly decreased number of non-zero entries (see (62)), while for $\cos\Omega_i \to -1$, the rank increases because the number of non-zero entries in (66) increases [21] (see (63)).

These statements can be directly extended to the diversity, since the $div(\cdot)$ diversity function of $\mathcal{S}\left(\mathcal{R}_\phi\left(\mathbf{T}(\mathcal{N})\right)\right)$ is evaluated via the number of non-zero entries in $\mathcal{S}\left(\mathcal{R}_\phi\left(\mathbf{T}(\mathcal{N})\right)\right)$,

$$div\big(\mathcal{S}\big(\mathcal{R}_\phi\big(\mathbf{T}(\mathcal{N})\big)\big)\big) = \bigcup_{\forall i,k} E_{i,k}\big(\mathcal{S}\big(\mathcal{R}_\phi\big(\mathbf{T}(\mathcal{N})\big)\big)\big) \neq 0, \tag{67}$$

where $E_{i,k}$ identifies an $(i,k)$ entry of $\mathcal{S}\big(\mathcal{R}_\phi\big(\mathbf{T}(\mathcal{N})\big)\big)$. Precisely, from (67) follows that $div\big(\mathcal{S}\big(\mathcal{R}_\phi\big(\mathbf{T}(\mathcal{N})\big)\big)\big)$ increases with the number $l$ of Gaussian sub-channels.

∎

## 4 Subcarrier Domain of Multiuser Multicarrier CVQKD

**Lemma 1** (Subcarrier domain of multiple-access multicarrier CVQKD). *The $\mathcal{R}_\phi\big(\mathbf{T}(\mathcal{N})\big)$ of $\mathbf{T}(\mathcal{N})$ in a $K_{in}, K_{out}$ multiuser setting is $\mathcal{R}_\phi^{K_{in},K_{out}}\big(\mathbf{T}(\mathcal{N})\big) = U_{K_{out}}\mathbf{T}(\mathcal{N})U_{K_{out}}$, where $U_{K_{out}}$ is a $K_{out} \times K_{out}$ unitary $U_{K_{out}} = \frac{1}{\sqrt{K_{out}}}e^{\frac{-i2\pi ik}{K_{out}}}$, $i,k = 0,...,K_{out}-1$.*

*Proof.*

Let $K_{in}, K_{out}$ be the number of transmitter and receiver users in a multiple access multicarrier CVQKD [3], and let $\mathbf{Z}$ be the $K_{in}$ dimensional input of the $K_{in}$ users. The Gaussian CV subcarriers formulate the $K_{in}$ dimensional vector

$$\mathbf{D} = U_{K_{in}}\mathbf{Z}, \tag{68}$$

where $U_{K_{in}}$ stands for the inverse CVQFT unitary operation.

The $U_{K_{in}}$ and $U_{K_{out}}$, $K_{in} \times K_{in}$, $K_{out} \times K_{out}$ unitary matrices at $l$ Gaussian sub-channels are as follows:

$$U_{K_{in}} = \frac{1}{\sqrt{K_{in}}}e^{\frac{i2\pi ik}{K_{in}}}, \; i,k = 0,...,K_{in}-1, \tag{69}$$

and

$$U_{K_{out}} = \frac{1}{\sqrt{K_{out}}}e^{\frac{-i2\pi ik}{K_{out}}}, \; i,k = 0,...,K_{out}-1, \tag{70}$$

which unitary is the CVQFT operation. (For further details, see the properties of the multicarrier CVQKD modulation in [2] and [3–6].)

Specifically, the output $\mathbf{Y}$ in a $K_{in}, K_{out}$ setting is then yielded as

$$\begin{aligned}\mathbf{Y} &= U_{K_{out}}\mathbf{T}(\mathcal{N})\big(U_{K_{out}}\mathbf{D}\big) + U_{K_{out}}\Delta \\ &= \big(U_{K_{out}}\mathbf{T}(\mathcal{N})U_{K_{out}}\big)\mathbf{D} + U_{K_{out}}\Delta \\ &= \mathcal{R}_\phi^{K_{in},K_{out}}\big(\mathbf{T}(\mathcal{N})\big)\mathbf{D} + U_{K_{out}}\Delta,\end{aligned} \tag{71}$$

thus without loss of generality

$$\mathcal{R}_\phi^{K_{in},K_{out}}\big(\mathbf{T}(\mathcal{N})\big) = U_{K_{out}}\mathbf{T}(\mathcal{N})U_{K_{out}}. \tag{72}$$

Particularly, by rewriting (52), $\mathcal{R}_\phi^{K_{in},K_{out}}\big(\mathbf{T}(\mathcal{N})\big)$ can be expressed as

$$\mathcal{R}_\phi^{K_{in},K_{out}}\left(\mathbf{T}\left(\mathcal{N}\right)\right)=$$
$$=\sum_{i=0}^{K_{in}-1}\sum_{k=0}^{K_{out}-1}b_{K_{out}}\left(\tfrac{k}{l}\right)^\dagger T_i\left(\mathcal{N}_i\right)b_{K_{in}}\left(\tfrac{i}{l}\right)$$
$$=\sum_{i=0}^{K_{in}-1}\sum_{k=0}^{K_{out}-1}b_{K_{out}}\left(\tfrac{k}{l}\right)^\dagger \mathrm{A}\left(\mathcal{N}_i\right)b_{K_{out}}\left(\cos\theta_i\right)b_{K_{in}}\left(\cos\theta_i^*\right)^\dagger b_{K_{in}}\left(\tfrac{i}{l}\right) \quad (73)$$
$$=\sum_{i=0}^{K_{in}-1}\sum_{k=0}^{K_{out}-1}x_i b_{K_{out}}\left(\tfrac{k}{l}\right)^\dagger b_{K_{out}}\left(\cos\theta_i\right)b_{K_{in}}\left(\cos\theta_i^*\right)^\dagger b_{K_{in}}\left(\tfrac{i}{l}\right),$$

where the basis vectors are precisely as

$$b_{K_{out}}\left(\cos\theta_i\right)=\frac{1}{\sqrt{K_{out}}}\begin{pmatrix}1\\ e^{\frac{-i2\pi l\cos\theta_i}{K_{out}}}\\ e^{\frac{-i2\pi 2l\cos\theta_i}{K_{out}}}\\ \vdots\\ e^{\frac{-i2\pi(K_{out}-1)l\cos\theta_i}{K_{out}}}\end{pmatrix},\text{ and }b_{K_{in}}\left(\cos\theta_i^*\right)=\frac{1}{\sqrt{K_{in}}}\begin{pmatrix}1\\ e^{\frac{-i2\pi l\cos\theta_i^*}{K_{in}}}\\ e^{\frac{-i2\pi 2l\cos\theta_i^*}{K_{in}}}\\ \vdots\\ e^{\frac{-i2\pi(K_{in}-1)l\cos\theta_i^*}{K_{in}}}\end{pmatrix}, \quad (74)$$

thus

$$b_{K_{out}}\left(\tfrac{k}{l}\right)=\frac{1}{\sqrt{K_{out}}}\begin{pmatrix}1\\ e^{\frac{-i2\pi k}{K_{out}}}\\ e^{\frac{-i2\pi 2k}{K_{out}}}\\ \vdots\\ e^{\frac{-i2\pi(K_{out}-1)k}{K_{out}}}\end{pmatrix},\text{ and }b_{K_{in}}\left(\tfrac{i}{l}\right)=\frac{1}{\sqrt{K_{in}}}\begin{pmatrix}1\\ e^{\frac{-i2\pi i}{K_{in}}}\\ e^{\frac{-i2\pi i 2}{K_{in}}}\\ \vdots\\ e^{\frac{-i2\pi i(K_{in}-1)}{K_{in}}}\end{pmatrix}. \quad (75)$$

Without loss of generality, the function $f$ from (36) can be rewritten as

$$f^{K_{out}}\left(\tau_i\right)=\frac{1}{K_{out}}e^{\frac{i\pi l(K_{out}-1)\tau_i}{K_{out}}}\frac{\sin\left(\pi l\tau_i\right)}{\sin\left(\pi\frac{l}{K_{out}}\tau_i\right)}, \quad (76)$$

with

$$\left|\cos\Omega_i\right|=\left|\frac{\sin\left(\pi l\left(\cos\theta_i-\cos\theta_i^*\right)\right)}{K_{out}\sin\left(\pi\frac{l}{K_{out}}\left(\cos\theta_i-\cos\theta_i^*\right)\right)}\right|, \quad (77)$$

and

$$f^{K_{out}}\left(\tfrac{k}{l}\right)=0,\text{ and }f^{K_{out}}\left(\tfrac{-k}{l}\right)=f^{K_{out}}\left(\tfrac{K_{out}-k}{l}\right),\ k=1,\ldots,K_{out}-1. \quad (78)$$

The maximum values of $f^{K_{out}}\left(\tfrac{k}{l}\right)$ are obtained at

$$\cos\theta_i=k/l\mod K_{out}/l. \quad (79)$$

The subcarrier domain representation $\mathcal{R}_\phi^{K_{in},K_{out}}\left(T_i\left(\mathcal{N}_i\right)\right)$ via $\left|f^{K_{out}}\left(\tau_i\right)\right|$ for $K_{out}>l$, at $\theta_i^*=\pi/2$, $l=2$ is shown in Fig. 5.

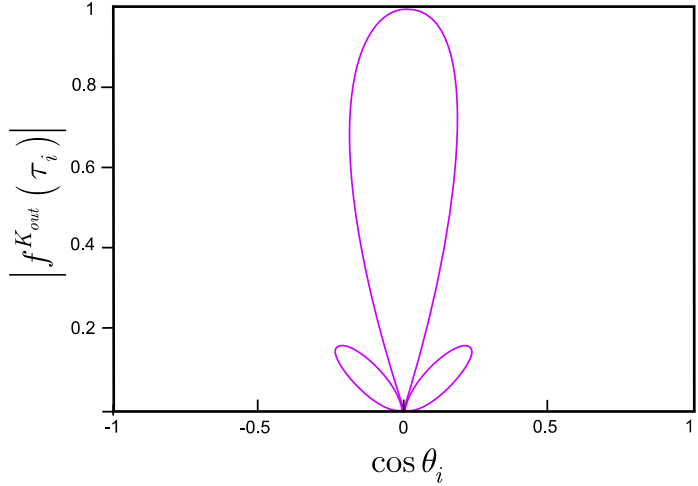

**Figure 5**. The function $\left|f^{K_{out}}(\tau_i)\right|$ of $\mathcal{R}_\phi^{K_{in},K_{out}}(T_i(\mathcal{N}_i))$ for $K_{out} > l$, at $\theta_i^* = \pi/2$, $l = 2$.

The sets $\mathrm{S}_{b_{K_{in}}}$, $\mathrm{S}_{b_{K_{out}}}$ of the $b_{K_{in}}$, $b_{K_{out}}$ orthonormal bases of a $K_{in}, K_{out}$ setting are as follows:

$$\mathrm{S}_{b_{K_{in}}} = \left\{b_{K_{in}}(0), b_{K_{in}}\left(\tfrac{1}{l}\right), \ldots, b_{K_{in}}\left(\tfrac{K_{in}-1}{l}\right)\right\} \in \mathcal{C}^{K_{in}}, \tag{80}$$

and

$$\mathrm{S}_{b_{K_{out}}} = \left\{b_{K_{out}}(0), b_{K_{out}}\left(\tfrac{1}{l}\right), \ldots, b_{K_{out}}\left(\tfrac{K_{out}-1}{l}\right)\right\} \in \mathcal{C}^{K_{out}}. \tag{81}$$

∎

# 5 Conclusions

We defined the subcarrier domain for multicarrier CVQKD. In a multicarrier CVQKD protocol, the characterization of the subcarrier domain of a Gaussian sub-channel is provided by the unitary CVQFT transformation, which has a central role in multicarrier CVQKD. The subcarrier domain injects physical attributes to the mathematical model of the Gaussian sub-channels. It provides a natural representation of multicarrier CVQKD and allows us to extend it to a multiple-access multicarrier CVQKD setting. The subcarrier domain representation is a general framework that can be utilized for an arbitrary multicarrier CVQKD scenario. The subcarrier domain also offers an apparatus to formulate the psychical model of the sub-channel transmission, which is particularly convenient for an experimental multicarrier CVQKD scenario.

# Acknowledgements

The author would like to thank Professor Sandor Imre for useful discussions. This work was partially supported by the GOP-1.1.1-11-2012-0092 (*Secure quantum key distribution between two units on optical fiber network*) project sponsored by the EU and European Structural Fund, and by the COST Action MP1006.

# Supplemental Information

## S.1 Notations

The notations of the manuscript are summarized in Table S.1.

**Table S.1.** Summary of notations.

| Symbol | Description |
|---|---|
| $rank(\cdot)$ | Rank function. |
| $div(\cdot)$ | Diversity function. |
| $i$ | Index for the $i$-th subcarrier Gaussian CV, $\lvert\phi_i\rangle = x_i + \mathrm{i}p_i$. |
| $j$ | Index for the $j$-th Gaussian single-carrier CV, $\lvert\varphi_j\rangle = x_j + \mathrm{i}p_j$. |
| $l$ | Number of Gaussian sub-channels $\mathcal{N}_i$ for the transmission of the Gaussian subcarriers. The overall number of the sub-channels is $n$. The remaining $n-l$ sub-channels do not transmit valuable information. |
| $x_i, p_i$ | Position and momentum quadratures of the $i$-th Gaussian subcarrier, $\lvert\phi_i\rangle = x_i + \mathrm{i}p_i$. |
| $x'_i, p'_i$ | Noisy position and momentum quadratures of Bob's $i$-th noisy subcarrier Gaussian CV, $\lvert\phi'_i\rangle = x'_i + \mathrm{i}p'_i$. |
| $x_j, p_j$ | Position and momentum quadratures of the $j$-th Gaussian single-carrier $\lvert\varphi_j\rangle = x_j + \mathrm{i}p_j$. |
| $x'_j, p'_j$ | Noisy position and momentum quadratures of Bob's $j$-th recovered single-carrier Gaussian CV $\lvert\varphi'_j\rangle = x'_j + \mathrm{i}p'_j$. |
| $x_{A,i}, p_{A,i}$ | Alice's quadratures in the transmission of the $i$-th subcarrier. |
| $\mathcal{R}_{\phi_i}(T_i(\mathcal{N}_i))$ | The subcarrier domain representation of sub-channel $\mathcal{N}_i$, $\mathcal{R}_{\phi_i}(T_i(\mathcal{N}_i)) = UT_i(\mathcal{N}_i)U$, where $U$ is the CVQFT unitary operation. |
| $\lvert\phi_i\rangle, \lvert\phi'_i\rangle$ | Transmitted and received Gaussian subcarriers. The subcarriers have angles $\theta_i^* \in [0, 2\pi]$, $\theta_i \in [0, 2\pi]$ CVs in the phase space $\mathcal{S}$. |
| $\mathbf{y}^{\mathcal{R}_\phi}$ | The subcarrier domain representation of output $\mathbf{y}$, expressed as $\mathbf{y}^{\mathcal{R}_\phi} = \sum_l \mathcal{R}_{\phi_i}(T_i(\mathcal{N}_i))d_i + F(\Delta_i)$. |
| A | The virtual gain of sub-channel $\mathcal{N}_i$, $\mathrm{A}(\mathcal{N}_i) = x_i$, where $x_i$ is a real variable. |
| $b(\cdot)$ | A basis vector of $\mathcal{R}_{\phi_i}$, evaluated as $\mathcal{R}_{\phi_i}(T_i(\mathcal{N}_i)) = \sum_k \mathrm{A}(\mathcal{N}_i) b(k/l)^\dagger b(\cos\theta_i) b(\cos\theta_i^*)^\dagger b(i/l)$, $k = 0 \ldots l-1$. |
| $\tau_i$ | The difference of the cos of phase space angles of the received and transmitted subcarriers, $\tau_i = \cos\theta_i - \cos\theta_i^*$. |
| $\lvert\cos\Omega_i\rvert$ | The cos of $\Omega_i$, where $\Omega_i = \theta_i - \theta_i^*$ is the angle of the basis vectors $b(\cos\theta_i)$, $b(\cos\theta_i^*)$. Defined as $\lvert\cos\Omega_i\rvert = \lvert b(\cos\theta_i^*)^\dagger b(\cos\theta_i)\rvert$, $\lvert\cos\Omega_i\rvert = \lvert f(\tau_i)\rvert$, and $\lvert\cos\Omega_i\rvert = \left\lvert\frac{\sin(\pi l(\cos\theta_i - \cos\theta_i^*))}{l\sin(\pi(\cos\theta_i - \cos\theta_i^*))}\right\rvert$. |
| $f(\tau_i)$ | Defines $\cos\Omega_i$, where $\Omega_i$ is the angle of the basis vectors $b(\cos\theta_i)$, $b(\cos\theta_i^*)$ expressed as $f(\tau_i) = \frac{1}{l}e^{i\pi(l-1)(\cos\theta_i - \cos\theta_i^*)}\frac{\sin(\pi l(\cos\theta_i - \cos\theta_i^*))}{\sin(\pi(\cos\theta_i - \cos\theta_i^*))}$. |
| $r$ | The period of function $\lvert f(\tau_i)\rvert$. |
| p | The plot of p: $(\cos\theta_i, \lvert f(\tau_i)\rvert)$. |
| $\mathrm{S}_b$ | The set $\mathrm{S}_b$ of the $\mathcal{R}_\phi$ orthonormal basis over the $\mathcal{C}^l$ complex space of the $\mathcal{R}_\phi$ subcarrier domain representation, $\mathrm{S}_b = \{b(0), b(\frac{1}{l}), \ldots, b(\frac{l-1}{l})\} \in \mathcal{C}^l$. |
| $\mathcal{G}_0$ | Set of subcarrier domain representations, $\mathcal{G}_0 = \{\lvert\mathcal{R}_{\phi_j}(T_j(\mathcal{N}_j))\rvert, j = 0, \ldots, s_0 - 1\} \subseteq \partial$, where $\partial = \{\lvert\mathcal{R}_{\phi_i}(T_i(\mathcal{N}_i))\rvert, i = 0, \ldots, l-1\}$. |
| $E_{i,k}(\mathbf{M})$ | The $(i,k)$ entry of matrix $\mathbf{M}$. |
| $U_{K_{out}}$ | The unitary CVQFT operation, $U_{K_{out}} = \frac{1}{\sqrt{K_{out}}} e^{\frac{-\mathrm{i}2\pi ik}{K_{out}}}$, $i,k = 0,\ldots,K_{out}-1$, $K_{out} \times K_{out}$ unitary matrix. |
| $U_{K_{in}}$ | The unitary inverse CVQFT operation, $U_{K_{in}} = \frac{1}{\sqrt{K_{in}}} e^{\frac{\mathrm{i}2\pi ik}{K_{in}}}$, $i,k = 0,\ldots,K_{in}-1$, $K_{in} \times K_{in}$ unitary matrix. |
| $\mathcal{R}_\phi^{K_{in},K_{out}}(\mathbf{T}(\mathcal{N}))$ | The subcarrier domain representation of $\mathbf{T}(\mathcal{N})$, expressed as $\mathcal{R}_\phi^{K_{in},K_{out}}(\mathbf{T}(\mathcal{N})) = U_{K_{out}} \mathbf{T}(\mathcal{N}) U_{K_{out}}$. |
| $b_{K_{out}}(\cdot)$, $b_{K_{in}}(\cdot)$ | The basis vectors of the subcarrier domain representation in a $K_{in}, K_{out}$ multiple-access multicarrier CVQKD scenario. |
| $f^{K_{out}}(\tau_i)$ | Defines $\cos\Omega_i$, where $\Omega_i$ is the angle of the basis vectors $b_{K_{out}}(\cos\theta_i)$, $b_{K_{in}}(\cos\theta_i^*)$ expressed as the angle $f^{K_{out}}(\tau_i) = \frac{1}{K_{out}} e^{\frac{\mathrm{i}\pi l(K_{out}-1)\tau_i}{K_{out}}} \frac{\sin(\pi l \tau_i)}{\sin\left(\pi \frac{l}{K_{out}} \tau_i\right)}$. |
| $S_{b_{K_{in}}}$, $S_{b_{K_{out}}}$ | The orthonormal bases of a $K_{in}, K_{out}$ setting, $S_{b_{K_{in}}} = \left\{ b_{K_{in}}(0), b_{K_{in}}\left(\frac{1}{l}\right), \ldots, b_{K_{in}}\left(\frac{K_{in}-1}{l}\right) \right\} \in \mathcal{C}^{K_{in}}$, $S_{b_{K_{out}}} = \left\{ b_{K_{out}}(0), b_{K_{out}}\left(\frac{1}{l}\right), \ldots, b_{K_{out}}\left(\frac{K_{out}-1}{l}\right) \right\} \in \mathcal{C}^{K_{out}}$. |
| $z \in \mathcal{CN}(0,\sigma_z^2)$ | The variable of a single-carrier Gaussian CV state, $\lvert\varphi_i\rangle \in \mathcal{S}$. Zero-mean, circular symmetric complex Gaussian random variable, $\sigma_z^2 = \mathbb{E}\left[\lvert z\rvert^2\right] = 2\sigma_{\omega_0}^2$, with i.i.d. zero mean, Gaussian random quadrature components $x, p \in \mathbb{N}(0, \sigma_{\omega_0}^2)$, where $\sigma_{\omega_0}^2$ is the variance. |
| $\Delta \in \mathcal{CN}(0,\sigma_\Delta^2)$ | The noise variable of the Gaussian channel $\mathcal{N}$, with i.i.d. zero-mean, Gaussian random noise components on the position and momentum quadratures $\Delta_x, \Delta_p \in \mathbb{N}(0, \sigma_\mathcal{N}^2)$, $\sigma_\Delta^2 = \mathbb{E}\left[\lvert \Delta \rvert^2\right] = 2\sigma_\mathcal{N}^2$. |
| $d \in \mathcal{CN}(0,\sigma_d^2)$ | The variable of a Gaussian subcarrier CV state, $\lvert\phi_i\rangle \in \mathcal{S}$. Zero-mean, circular symmetric Gaussian random variable, |
| | $\sigma_d^2 = \mathbb{E}\left[\left|d\right|^2\right] = 2\sigma_\omega^2$, with i.i.d. zero mean, Gaussian random quadrature components $x_d, p_d \in \mathbb{N}\left(0,\sigma_\omega^2\right)$, where $\sigma_\omega^2$ is the modulation variance of the Gaussian subcarrier CV state. |
| $F^{-1}(\cdot) = \text{CVQFT}^\dagger(\cdot)$ | The inverse CVQFT transformation, applied by the encoder, continuous-variable unitary operation. |
| $F(\cdot) = \text{CVQFT}(\cdot)$ | The CVQFT transformation, applied by the decoder, continuous-variable unitary operation. |
| $F^{-1}(\cdot) = \text{IFFT}(\cdot)$ | Inverse FFT transform, applied by the encoder. |
| $\sigma_{\omega_0}^2$ | Single-carrier modulation variance. |
| $\sigma_\omega^2 = \frac{1}{l}\sum_l \sigma_{\omega_i}^2$ | Multicarrier modulation variance. Average modulation variance of the $l$ Gaussian sub-channels $\mathcal{N}_i$. |
| $\left|\phi_i\right\rangle = \left|\text{IFFT}\left(z_{k,i}\right)\right\rangle$ $= \left|F^{-1}\left(z_{k,i}\right)\right\rangle = \left|d_i\right\rangle.$ | The $i$-th Gaussian subcarrier CV of user $U_k$, where IFFT stands for the Inverse Fast Fourier Transform, $\left|\phi_i\right\rangle \in \mathcal{S}$, $d_i \in \mathcal{CN}\left(0,\sigma_{d_i}^2\right)$, $\sigma_{d_i}^2 = \mathbb{E}\left[\left|d_i\right|^2\right]$, $d_i = x_{d_i} + ip_{d_i}$, $x_{d_i} \in \mathbb{N}\left(0,\sigma_{\omega_F}^2\right)$, $p_{d_i} \in \mathbb{N}\left(0,\sigma_{\omega_F}^2\right)$ are i.i.d. zero-mean Gaussian random quadrature components, and $\sigma_{\omega_F}^2$ is the variance of the Fourier transformed Gaussian state. |
| $\left|\varphi_{k,i}\right\rangle = \text{CVQFT}\left(\left|\phi_i\right\rangle\right)$ | The decoded single-carrier CV of user $U_k$ from the subcarrier CV, expressed as $F\left(\left|d_i\right\rangle\right) = \left|F\left(F^{-1}\left(z_{k,i}\right)\right)\right\rangle = \left|z_{k,i}\right\rangle$. |
| $\mathcal{N}$ | Gaussian quantum channel. |
| $\mathcal{N}_i, i = 1,\ldots,n$ | Gaussian sub-channels. |
| $T(\mathcal{N})$ | Channel transmittance, normalized complex random variable, $T(\mathcal{N}) = \text{Re}\, T(\mathcal{N}) + i\,\text{Im}\, T(\mathcal{N}) \in \mathcal{C}$. The real part identifies the position quadrature transmission, the imaginary part identifies the transmittance of the position quadrature. |
| $T_i(\mathcal{N}_i)$ | Transmittance coefficient of Gaussian sub-channel $\mathcal{N}_i$, $T_i(\mathcal{N}_i) = \text{Re}\left(T_i(\mathcal{N}_i)\right) + i\,\text{Im}\left(T_i(\mathcal{N}_i)\right) \in \mathcal{C}$, quantifies the position and momentum quadrature transmission, with |
| | (normalized) real and imaginary parts $0 \leq \operatorname{Re} T_i(\mathcal{N}_i) \leq 1/\sqrt{2}$, $0 \leq \operatorname{Im} T_i(\mathcal{N}_i) \leq 1/\sqrt{2}$, where $\operatorname{Re} T_i(\mathcal{N}_i) = \operatorname{Im} T_i(\mathcal{N}_i)$. |
| $T_{Eve}$ | Eve's transmittance, $T_{Eve} = 1 - T(\mathcal{N})$. |
| $T_{Eve,i}$ | Eve's transmittance for the $i$-th subcarrier CV. |
| $\mathbf{z} = \mathbf{x} + \mathrm{i}\mathbf{p} = (z_1,...,z_d)^T$ | A $d$-dimensional, zero-mean, circular symmetric complex random Gaussian vector that models $d$ Gaussian CV input states, $\mathcal{CN}(0, \mathbf{K_z})$, $\mathbf{K_z} = \mathbb{E}[\mathbf{zz}^\dagger]$, where $z_i = x_i + \mathrm{i}p_i$, $\mathbf{x} = (x_1,...,x_d)^T$, $\mathbf{p} = (p_1,...,p_d)^T$, with $x_i \in \mathbb{N}(0, \sigma^2_{\omega_0})$, $p_i \in \mathbb{N}(0, \sigma^2_{\omega_0})$ i.i.d. zero-mean Gaussian random variables. |
| $\mathbf{d} = F^{-1}(\mathbf{z})$ | An $l$-dimensional, zero-mean, circular symmetric complex random Gaussian vector of the $l$ Gaussian subcarrier CVs, $\mathcal{CN}(0, \mathbf{K_d})$, $\mathbf{K_d} = \mathbb{E}[\mathbf{dd}^\dagger]$, $\mathbf{d} = (d_1,...,d_l)^T$, $d_i = x_i + \mathrm{i}p_i$, $x_i, p_i \in \mathbb{N}(0, \sigma^2_{\omega_F})$ are i.i.d. zero-mean Gaussian random variables, $\sigma^2_{\omega_F} = 1/\sigma^2_{\omega_0}$. The $i$-th component is $d_i \in \mathcal{CN}(0, \sigma^2_{d_i})$, $\sigma^2_{d_i} = \mathbb{E}[|d_i|^2]$. |
| $\mathbf{y}_k \in \mathcal{CN}(0, \mathbb{E}[\mathbf{y}_k\mathbf{y}_k^\dagger])$ | A $d$-dimensional zero-mean, circular symmetric complex Gaussian random vector. |
| $y_{k,m}$ | The $m$-th element of the $k$-th user's vector $\mathbf{y}_k$, expressed as $y_{k,m} = \sum_l F(T_i(\mathcal{N}_i))F(d_i) + F(\Delta_i)$. |
| $F(\mathbf{T}(\mathcal{N}))$ | Fourier transform of $\mathbf{T}(\mathcal{N}) = [T_1(\mathcal{N}_1)...,T_l(\mathcal{N}_l)]^T \in \mathcal{C}^l$, the complex transmittance vector. |
| $F(\Delta)$ | Complex vector, expressed as $F(\Delta) = e^{\frac{-F(\Delta)^T \mathbf{K}_{F(\Delta)} F(\Delta)}{2}}$, with covariance matrix $\mathbf{K}_{F(\Delta)} = \mathbb{E}[F(\Delta)F(\Delta)^\dagger]$. |
| $\mathbf{y}[j]$ | AMQD block, $\mathbf{y}[j] = F(\mathbf{T}(\mathcal{N}))F(\mathbf{d})[j] + F(\Delta)[j]$. |
| $\tau = \|F(\mathbf{d})[j]\|^2$ | An exponentially distributed variable, with density $f(\tau) = (1/2\sigma^{2n}_\omega)e^{-\tau/2\sigma^2_\omega}$, $\mathbb{E}[\tau] \leq n2\sigma^2_\omega$. |
| $T_{Eve,i}$ | Eve's transmittance on the Gaussian sub-channel $\mathcal{N}_i$, $T_{Eve,i} = \operatorname{Re} T_{Eve,i} + \mathrm{i} \operatorname{Im} T_{Eve,i} \in \mathcal{C}$, $0 \leq \operatorname{Re} T_{Eve,i} \leq 1/\sqrt{2}$, $0 \leq \operatorname{Im} T_{Eve,i} \leq 1/\sqrt{2}$, $0 \leq \left|T_{Eve,i}\right|^2 < 1$. |
| $d_i$ | A $d_i$ subcarrier in an AMQD block. |
| $\nu_{\min}$ | The $\min\{\nu_1,\ldots,\nu_l\}$ minimum of the $\nu_i$ sub-channel coefficients, where $\nu_i = \sigma_{\mathcal{N}}^2 \big/ \left|F\left(T_i\left(\mathcal{N}_i\right)\right)\right|^2$ and $\nu_i < \nu_{Eve}$. |
| $\sigma_\omega^2$ | Modulation variance, $\sigma_\omega^2 = \nu_{Eve} - \nu_{\min} \mathcal{G}(\delta)_{p(x)}$, where $\nu_{Eve} = \frac{1}{\lambda}$, $\lambda = \left|F\left(T_{\mathcal{N}}^*\right)\right|^2 = \frac{1}{n}\sum_{i=0}^{n-1}\left|\sum_{k=0}^{n-1} T_k^* e^{\frac{-\mathrm{i}2\pi ik}{n}}\right|^2$ and $T_{\mathcal{N}}^*$ is the expected transmittance of the Gaussian sub-channels under an optimal Gaussian collective attack. |
| $\nu_\kappa$ | Additional sub-channel coefficient for the correction of modulation imperfections. For an ideal Gaussian modulation, $\nu_\kappa = 0$, while for an arbitrary $p(x)$ distribution $\nu_\kappa = \nu_{\min}\left(1 - \mathcal{G}(\delta)_{p(x)}\right)$, where $\kappa = \frac{1}{\nu_{Eve} - \nu_{\min}\left(\mathcal{G}(\delta)_{p(x)} - 1\right)}$. |
| $\sigma_{\omega_i'}^2$ | The constant modulation variance $\sigma_{\omega_i'}^2$ for eigenchannel $\lambda_i$, evaluated as $\sigma_{\omega_i'}^2 = \mu - \left(\sigma_{\mathcal{N}}^2 \big/ \max_{n_{\min}} \lambda_i^2\right) = \frac{1}{n_{\min}}\sigma_{\omega'}^2$, with a total constraint $\sigma_{\omega'}^2 = \sum_{n_{\min}} \sigma_{\omega_i'}^2 = \frac{1}{l}\sum_l \sigma_{\omega_i}^2 = \sigma_\omega^2$. |
| $\sigma_{\omega''}^2$ | The modulation variance of the AMQD multicarrier transmission in the SVD environment. Expressed as $\sigma_{\omega''}^2 = \nu_{Eve} - \left(\sigma_{\mathcal{N}}^2 \big/ \max_{n_{\min}} \lambda_i^2\right)$, where $\lambda_i$ is the $i$-th eigenchannel of $F(\mathbf{T})$, $\max_{n_{\min}} \lambda_i^2$ is the largest eigenvalue of $F(\mathbf{T})F(\mathbf{T})^\dagger$, with a total constraint $\frac{1}{l}\sum_l \sigma_{\omega_i''}^2 = \sigma_{\omega''}^2 > \sigma_\omega^2$. |
| $\mathcal{S}(F(\mathbf{T}))$ | A statistical model of $F(\mathbf{T})$. |
| $\mathcal{S}\left(\mathcal{R}_{\phi_i}\left(T_i\left(\mathcal{N}_i\right)\right)\right)$ | A statistical model of $\mathcal{R}_{\phi_i}\left(T_i\left(\mathcal{N}_i\right)\right)$. |

## S.2 Abbreviations

| | |
|---|---|
| AMQD | Adaptive Multicarrier Quadrature Division |
| CV | Continuous-Variable |
| CVQFT | Continuous-Variable Quantum Fourier Transform |
| CVQKD | Continuous-Variable Quantum Key Distribution |
| DV | Discrete Variable |
| FFT | Fast Fourier Transform |
| IFFT | Inverse Fast Fourier Transform |
| MQA | Multiuser Quadrature Allocation |
| QKD | Quantum Key Distribution |
| SNR | Signal to Noise Ratio |
| SVD | Singular Value Decomposition |